\documentclass[aps,pra,twocolumn,superscriptaddress,floatfix,nofootinbib,showpacs,longbibliography,groupedaddress]{revtex4-1}

\usepackage{xcolor}
\usepackage[utf8]{inputenc}  
\usepackage[T1]{fontenc}     
\usepackage[british]{babel}  
\usepackage[sc,osf]{mathpazo}
\usepackage[scaled=0.86]{berasans}  
\usepackage[colorlinks=true, citecolor=blue, urlcolor=blue]{hyperref}  
\usepackage{graphicx} 
\usepackage[babel]{microtype}  
\usepackage{amsmath,amssymb,amsthm,bm,amsfonts,mathrsfs,bbm} 

\usepackage{xspace}  
\usepackage{xcolor}
\usepackage{multirow}
\usepackage{array}
\usepackage{bigstrut}
\usepackage{braket}
\usepackage{color}
\usepackage{natbib}
\usepackage{multirow}
\usepackage{mathtools}
\usepackage{float}
\usepackage[caption = false]{subfig}
\usepackage{xcolor,colortbl}
\usepackage{color}

\newcommand{\be}{\begin{equation}}
\newcommand{\ee}{\end{equation}}
\newcommand{\ba}{\begin{eqnarray}}
\newcommand{\ea}{\end{eqnarray}}

\newtheorem{proposition}{Proposition}

\def\>{\rangle}
\def\<{\langle}

\begin{document}
\title{Bell Nonlocality and the Reality of Quantum Wavefunction}

\author{Anandamay Das Bhowmik}
\affiliation{Physics and Applied Mathematics Unit, Indian Statistical Institute, 203 BT Road, Kolkata, India.}

\author{Preeti Parashar}
\affiliation{Physics and Applied Mathematics Unit, Indian Statistical Institute, 203 BT Road, Kolkata, India.}

\author{Manik Banik}
\affiliation{School of Physics, IISER Thiruvananthapuram, Vithura, Kerala  695551, India.}

\begin{abstract}
Status of quantum wavefunction is one of the most debated issues in quantum foundations -- whether it corresponds directly to the reality or just represents knowledge or information about some aspect of reality. In this letter we propose a {\it $\psi$-ontology} theorem that excludes a class of ontological explanations where quantum wavefunction is treated as mere information. Our result, unlike the acclaimed Pusey-Barrett-Rudolph's theorem, does not presume the absence of \textit{holistic} ontological properties for product quantum preparations. At the core of our derivation we utilize the seminal no-go result by John S. Bell that rules out any {\it local realistic} world view for quantum theory. We show that the observed phenomenon of quantum nonlocality cannot be incorporated in a class of $\psi$-epistemic models. Using the well known Clauser-Horne-Shimony-Holt inequality we obtain a threshold bound on the degree of epistemicity above which the ontological models are not compatible with quantum statistics. 
\end{abstract}

\maketitle
\section{Introduction} 
What does a quantum wavefunction stand for? Does it represent the state of reality of the physical system ($\psi$-ontic doctrine) or it merely does provide information about the system ($\psi$-epistemic doctrine)? This question is at the core of quantum foundational debate ever since the advent of the theory \cite{Bacciagaluppi09}. A mathematically precise formulation of this question, within a broad class of realist approaches to quantum theory, can be made in the ontological model framework of Harrigan \& Spekkens \cite{Harrigan09} (see also \cite{Spekkens05,Harrigan07}). Epistemicity, in this framework, is defined (as well as quantified) through the amount of overlap between probability distributions over the ontic states resulting from different quantum preparations. 

In a fascinating development, Pusey, Barrett, and Rudolph (PBR) have shown that $\psi$-epistemic interpretation contradicts prediction of quantum theory in any model where independently prepared systems have independent physical states (named as preparation independence (PI) assumption) \cite{Pusey12}. This result drew attention of quantum foundations community and within few days several researchers reported similar theorems \cite{Hardy13,Colbeck12,Colbeck17,Patra13}, commonly called as $\psi$-ontology theorems \cite{Leifer14}, derived under different assumptions. Subsequently, several criticisms were raised regarding the assumptions used in those $\psi$-ontology theorems \cite{Schlosshauer12,Emerson13,Ballentine14,Schlosshauer14} (see also \cite{Leifer14}). In particular, the authors in \cite{Schlosshauer14} have shown that the physical rationale for composition principles such as PI overreaches and thus places the no-go theorem put forward by PBR into jeopardy.  

Interestingly, Owen Maroney came up with a new kind of $\psi$-ontology theorem that uses no compositional assumption and rules out a class of ontological models with certain degree of epistemicity \cite{Maroney12,Maroney12(1)}. Subsequently, several other results were obtained excluding $\psi$-epistemic models with increasingly lower degree of epistemicity and consequently imposing higher degree of onticity on quantum wavefunction \cite{Leifer13,Barrett14,Leifer14(1),Branciard14,Ringbauer15}. However, all these theorems apply to Hilbert spaces of dimension strictly greater than two. 

In this work we derive a new $\psi$-ontology theorem that excludes the maximally $\psi$-epistemic model as well as non-maximal ones with certain degree of epistemicity. Importantly our theorem does not assume the ontic composition principles of PI. In fact it does not presuppose that ontic state space for quantum product preparations should be Cartesian product of their individual ontic state spaces only. In other words, we consider that the two or more quantum systems prepared even in product state can possess {\it holistic} ontic features accessible only through global measurements, which broadens the scope of our result over PBR's theorem. Furthermore, unlike the $\psi$-ontology theorems reported in Refs.\cite{Maroney12,Maroney12(1),Leifer13,Barrett14,Leifer14(1),Branciard14,Ringbauer15}, the present theorem applies to qubit Hilbert space too. Quite importantly, our result demonstrates an interesting connection between degree of epistemicity and Bell nonlocality \cite{Bell64,Bell66,Mermin93,Brunner14}. While it was already recognized that $\psi$-complete and $\psi$-ontic models for quantum theory are inconsistent with the concept of {\it locality} \cite{Bacciagaluppi09,Harrigan09}, our result establishes the fact in the reverse direction. It shows that the phenomenon of Bell nonlocality prefers $\psi$-ontic interpretation for quantum wavefunction as $\psi$-epistemicity models having epistemicity above a threshold degree cannot incorporate the observed quantum nonlocality.     
  
\section{Framework} 
We first recall the ontological model framework as developed in \cite{Harrigan09} (see also \cite{Harrigan07,Spekkens05,Leifer14,Ballentine14}). Such a model consists of a space $\Lambda$ of ontic states that completely identify the possible physical properties of a system. A quantum system prepared in the state $\ket{\psi}$ is associated with a probability distribution $\mu(\lambda|\psi)$ over $\Lambda$, where each realization of the preparation $\ket{\psi}$ results in an ontic state $\lambda\in\Lambda$ sampled with probability measure $\mu(\lambda|\psi)$. The probability distribution of the resulting ontic states is called the epistemic state associated with $\ket{\psi}$ and $\mbox{Supp}[\mu(\lambda|\ket{\psi})]\equiv\Lambda_{\psi}:=\{\lambda\in\Lambda~|~\mu(\lambda|\psi)>0\}$ is called the ontic support of $\ket{\psi}$. For all $\ket{\psi}$, we must have $\int_{\Lambda_{\psi}}d\lambda\mu(\lambda|\psi)=1$. Whenever an observable $M$ is measured on a system, the possible outcomes are the eigenvalues $\phi_k$ with the associated eigenvectors $\ket{\phi_k}$, {\it i.e.}, $M=\sum_k \phi_k\ket{\phi_k}\bra{\phi_k}$. Here we restrict our attention to rank-1 projective measurements only, although quantum theory allows more general measurement process described by positive operator valued measure (POVM). Given a system in the ontic state $\lambda$, the probability of obtaining the $k^{th}$ outcome is given by an response function $\xi(\phi_k|\lambda, M)\in[0,1]$. A generic ontological model keeps open the possibility for this outcome response to be contextual \cite{Spekkens05}; whenever measurement context is not important we will denote the response function simply as $\xi(\phi_k|\lambda)$. An {\it outcome deterministic} (OD) ontological model demands $\xi(\phi_k|\lambda, M)\in\{0,1\}~\forall~k,\lambda,M$. Denoting $\mbox{Supp}[\xi(\psi|\lambda)]:=\{\lambda\in\Lambda~|~\xi(\psi|\lambda)>0\}$ and $\mbox{Core}[\xi(\psi|\lambda)]:=\{\lambda\in\Lambda~|~\xi(\psi|\lambda)=1\}$, the following set inclusion relations are immediate: $$\Lambda_\psi\subseteq\mbox{Core}[\xi(\psi|\lambda)]\subseteq\mbox{Supp}[\xi(\psi|\lambda)].$$
An OD model satisfies $\mbox{Core}[\xi(\psi|\lambda)]=\mbox{Supp}[\xi(\psi|\lambda)]$, whereas a model with $\Lambda_\psi=\mbox{Core}[\xi(\psi|\lambda)]$ is termed as {\it reciprocal} \cite{Ballentine14}. Interestingly, the author in \cite{Ballentine14} have also established that
\begin{equation}\label{eq1}
    Maximally~\psi-epistemic~\Leftrightarrow~OD~\wedge~Reciprocal.
\end{equation}
An operational transformation procedures $T$ at ontological level corresponds to a transition matrix $\Gamma_T(\lambda,\tilde{\lambda})$ denoting the probability density for a transition from the ontic state $\lambda$ to the ontic state $\tilde{\lambda}$. In prepare and measure scenario, the reproducibility of the Born rule at operational level demands $\int_{\Lambda}d\lambda\xi(\phi_k|\lambda,M)\mu(\lambda|\psi)=|\langle\phi_k|\psi\rangle|^2:=\mbox{Pr}(\phi_k|\psi)$. 

\section{Degree of epistemicity} 
In a maximally $\psi$-epistemic model the quantum overlap $|\langle\psi|\phi\rangle|^2$ for any two state vectors $\ket{\psi}$ and $\ket{\phi}$ is completely accounted for by the overlap between the corresponding epistemic distributions $\mu(\lambda|\psi)$ and $\mu(\lambda|\phi)$, {\it i.e.}, $\int_{\Lambda_\phi}\mu(\lambda|\psi)d\lambda=|\langle\psi|\phi\rangle|^2$ \cite{Maroney12}. Maroney's theorem \cite{Maroney12} and the subsequent results \cite{Maroney12(1),Leifer13,Barrett14,Leifer14(1),Branciard14,Ringbauer15} exclude maximally $\psi$-epistemic model and a class of ontological models with increasingly lower degree of epistemicity. To quantify the degree of epistemicity of an ontological model, please note that,$\int_{\Lambda_\phi}d\lambda\mu(\lambda|\psi)=\int_{\Lambda_\phi}d\lambda\xi(\phi|\lambda)\mu(\lambda|\psi)\le \int_{\Lambda}d\lambda\xi(\phi|\lambda)\mu(\lambda|\psi)=|\langle\phi|\psi\rangle|^2$. The first equality is due to the fact that $\Lambda_\phi\subseteq\mbox{Core}[\xi(\phi|\lambda)]$. One can express the above inequality as a equality of the following form
\begin{equation}\label{deg}
\int_{\Lambda_\phi}d\lambda\mu(\lambda|\psi)=\Omega(\phi,\psi)~|\langle\phi|\psi\rangle|^2,
\end{equation}
where $\Omega(\phi,\psi)$ captures degree of epistemicity of a model and for any pair of non-orthogonal quantum states $\Omega(\phi,\psi)\in[0,1]$. For a maximally $\psi$-ontic model $\Omega(\phi,\psi)=0$ for all non-orthogonal pairs of state, while $\Omega(\phi,\psi)=1$ for all such pairs in a maximally $\psi$-ontic model (see \cite{Maroney12} for more elaboration). Larger the departure of $\Omega(\phi,\psi)$ from its maximum possible value more the model is $\psi$-ontic and consequently less  $\psi$-epistemic it is.

\section{Ontic composition} 
The discussions so far consider ontological models for a single system only. More involving situations arise for composite systems. Consider System-$A$ and System-$B$ with ontic state space $\Lambda^A$ and $\Lambda^B$, respectively. For a composite quantum system prepared in a state $\rho_{AB}$, a naive classical thinking suggests the joint ontic state $\lambda_{joint}$ to be in $\Lambda_A\times\Lambda_B$, {\it i.e.}, $\Lambda_{\rho_{AB}}\subseteq\Lambda^A\times\Lambda^B$. However, lesson from the seminal Bell's theorem indicates towards a more intricate description of the ontic state for composite system. Violation of any local realistic inequality by some joint quantum preparation $\rho_{AB}$ necessitates some `nonlocal' ontic state space $\Lambda^{NL}$. To say more precisely, this nonlocal variable captures the essence of nonlocal correlation in the sense of Bell \cite{Bell64,Bell66}. At this point it should be reminded that these nonlocal correlations are perfectly compatible with no-signaling principle that prohibits instantaneous transfer of information. Therefore, we can say, $\Lambda_{\rho_{AB}}\subseteq(\Lambda^A\times\Lambda^B)\cup\Lambda^{NL}$, where $\Lambda^{NL}$ accounts for Bell type local realistic inequality violation. Importantly, all product states are Bell local (same is the case for separable state, and {\it local} entangled states \cite{Werner89,Barrett02,Bowles15}). But, does it assert that such a product quantum preparation should not contain any {\it holistic} ontic feature, {\it i.e} their ontic support ought to be a subset of $\Lambda^A\times\Lambda^B$? PBR in their derivation have considered such an assumption. In fact their PI assumption is even restrictive as it explicitly spells  $\mu(\lambda_A,\lambda_B|\psi_A\otimes\psi_B)=\mu(\lambda_A|\psi_A)\mu(\lambda_B|\psi_B)$ \cite{Pusey12}. However, such an assumption is conservative as it considers (local) reality  of the product measurements only. It is possible that a composite system even prepared in product state contains properties that are accessible through global measurements only, {\it e.g.}, Bell basis measurement. More dramatically, the phenomena of `nonlocality without entanglement' indeed indicates towards such a situation even without involving any entanglement in the measurement basis \cite{Bennett99,Halder19,Bhattacharya20}. Thus for a bipartite product (also for separable and {\it local} entangled) quantum state $\rho_{AB}$ the ontic support should be considered as $\Lambda_{\rho_{AB}}\subseteq\Lambda^A\times\Lambda^B\times\Lambda^{G}$, where $\Lambda^G$ carries strictly relational holistic information about the two systems and remains hidden under local measurements, {\it i.e.} presumed not to be accessible by local measurements. Here we make a clear distinction between $\Lambda^{G}$ and $\Lambda^{NL}$. Both are non-classical features of ontological state space for bipartite quantum system. But they correspond to two completely distinct operational non-classicality of quantum theory. $\Lambda^{NL}$ contains variables that are nonlocal strictly in the sense of Bell inequality violation, which is revealed through local measurements performed on spatially separated subsystems of the composite system - no global measurement is involved or required in this case. This is clearly not the case for ontic elements belonging to $\Lambda^{G}$, that accounts for the global properties corresponding to non-classical joint measurements allowed in quantum theory due to its richer bipartite effect space structure containing entangled effects in addition to product effects. These global measurements with entangled effects correspond to holistic properties or observables of the composite system whose outcomes can never be realised by locally measuring the individual subsystems. At this point one may consider the possibility that the product quantum states have ontic support within $\Lambda^{NL}$ in such a way the `nonlocal' effect gets averaged out in the operational statistics. However, our following proposition (see \cite{Supple} for the proof) puts a {\it no-go} on such assertion. 
\begin{proposition}\label{prop1}
In a maximally $\psi$-epistemic model (more generally in any outcome deterministic ontological model) quantum product preparations do not possess any nonlocal ontic state. 
\end{proposition}
At this point we leave this question open for further investigation weather the proposition can be extended for any $\psi$-epistemic model. However, for other $\psi$-epistemic models (apart from those considered in Proposition \ref{prop1}) nonexistence of nonlocal ontic state in the support of quantum product preparation can be justified from the principle of {\it Occam's razor}, which researchers in quantum foundation have applied in different context \cite{Ionicioiu11}. For an entangled state $\rho_{AB}$ exhibiting Bell nonlocality the ontic description will be $\Lambda_{\rho_{AB}}\subseteq(\Lambda^A\times\Lambda^B\times\Lambda^{G})\cup\Lambda^{NL}$ with $\Lambda_{\rho_{AB}}\cap\Lambda^{NL}\neq\emptyset$. The Cartesian product structure assumed above is not necessary for our argument but it provides a simple way to present the idea.

\section{A new $\psi$-ontology theorem} 
We are now in a position to prove our $\psi$-ontology theorem. To this aim we consider a quantum copying machine $\mathbb{M}$ that perfectly copies the states $\ket{0}$ and $\ket{1}$. The pioneering `no-cloning' theorem does not prohibit existence of such a machine as the coping states are mutually orthogonal \cite{Wootters82}. The action of $\mathbb{M}$ can be described by a unitary evolution $U_{\mathbb{M}}$ satisfying $U_{\mathbb{M}}\ket{i}\ket{r} = \ket{i}\ket{i}$, where $\ket{r}$ is some fixed reference state and $i\in\{0,1\}$. The machine has two input ports, one fed with particles $R$ prepared in some reference state $\ket{r}$ and the other fed with system $S$ prepared in state $\ket{0}$ or $\ket{1}$. 

The action of this copying machine is worth analyzing in the ontological picture. Whenever $S$ is prepared in the state $\ket{i}$, a composite ontic state $\lambda_{joint}= (\lambda^S,\lambda^R,\lambda^G)$ is fed into the machine, where $\lambda^S \in \Lambda_i\subseteq\Lambda^S,~\lambda^R\in \Lambda_r\subseteq\Lambda^R$, and $\lambda^G \in \Lambda^G$, {\it i.e.} $\Lambda_{ir}:=\{\mu(\lambda_{joint}|\ket{ir})>0\} \subseteq \Lambda^S \times \Lambda^R \times \Lambda^G$. After the successful completion of cloning, the machine yields the outcome $\ket{ii}$, {\it i.e.}, it yields an ontic state belonging in $\Lambda_{ii} \subseteq \Lambda^S \times \Lambda^R \times \Lambda^G $. Let us now feed the above copying machine with the system state prepared in $\ket{+}=(\ket{0}+\ket{1})/\sqrt{2}$. At the ontological level the machine receives some ontic state $\lambda_{joint}\in\Lambda^S \times \Lambda^R \times \Lambda^G$ sampled with probability distribution $\mu(\lambda_{joint}|\ket{+r})=\mu(\lambda^S,\lambda^R,\lambda^G|\ket{+r})$. Let us consider a maximally $\psi$-epistemic model underlying quantum theory. This will imply $\Lambda_+=(\Lambda_0\cap\Lambda_+)\cup(\Lambda_1\cap\Lambda_+)$ and $\Lambda_{+r}=(\Lambda_{0r}\cap\Lambda_{+r})\cup(\Lambda_{1r}\cap\Lambda_{+r})\subseteq\Lambda_{0r}\cup\Lambda_{1r}$. Since the ontic state obtained after the machine's action only  depends on the input ontic state fed into the machine, therefore $\mbox{Supp}[\mu(\lambda_{joint}|U_{\mathbb{M}}[\ket{+r}])]\subseteq\mbox{Supp}[\mu(\lambda_{joint}|U_{\mathbb{M}}[\ket{0r}])]\cup\mbox{Supp}[\mu(\lambda_{joint}|U_{\mathbb{M}}[\ket{+r}])]\equiv\Lambda_{00}\cup\Lambda_{11}$. Since a product preparation can't have any `nonlocal' ontic reality, {\it i.e.} $\Lambda_{00}\cap\Lambda^{NL}=\emptyset=\Lambda_{11}\cap\Lambda^{NL}$ (see Proposition \ref{prop1}), therefore it, in turn, implies $\mbox{Supp}[\mu(\lambda_{joint}|U_{\mathbb{M}}[\ket{+r}])]\cap\Lambda^{NL}=\emptyset$.

But the above conclusion is in direct contradiction with predictions of quantum theory. Due to linearity of the machine's action one will obtain the output $\ket{\phi^+}=(\ket{0}\ket{0}+\ket{1}\ket{1})/\sqrt{2}$ whenever the machine $\mathbb{M}$ is supposed to copy the state $\ket{+}$. Being the maximally entangled state it exhibits maximum violation of Clauser-Horne-Shimony-Holt (CHSH) inequality \cite{Clauser69} for suitable measurement choice on its local parts. Hence the entire ontic support of the output pairs cannot be contained within $\Lambda^{S} \times \Lambda^{R} \times \Lambda^G$; in other words $\Lambda_{\phi^+} \cap \Lambda^{NL}\neq \emptyset$. At this point it is important to note that, like PBR we have neither considered ontic state space for a product preparation to be Cartesian product of their individual state space nor we presume the {\it strong} assumption of `preparation independence'. We have not also invoked the assumption of `Local independence', {\it i.e.} $\int d\lambda^G \mu(\lambda^S, \lambda^R,\lambda^G|\ket{ir})= \mu(\lambda^S|i) \mu(\lambda^R|r)$ as considered in \cite{Emerson13} and the assumption of `Ontic indifference' used in \cite{Hardy13}.

A similar proof runs if we feed any state $\ket{\psi}=\alpha\ket{0}+\beta\ket{1}$ into the machine $\mathbb{M}$ instead of the state $\ket{+}$, where $\mathbb{C}\ni\alpha,\beta\neq 0$ and $|\alpha|^2+|\beta|^2=1$. In a maximally $\psi$-epistemic theory the ontic support $\Lambda_\psi$ entirely gets shared between $\Lambda_0$ and $\Lambda_1$. Whereas for $\ket{+}$ these two shares are equal, for $\ket{\psi}$ the shares are proportional to the quantum overlaps. Similar argument in this case implies $\mbox{Supp}[\mu(\lambda_{joint}|U_{\mathbb{M}}[\ket{\psi r}])]\cap\Lambda^{NL}=\emptyset$. But this is again in contradiction with quantum prediction as the machine's action on $\ket{\psi r}$ yields the state $\alpha\ket{00}+\beta\ket{11}$ which is known to be Bell nonlocal \cite{Gisin91}.

Our result thus establishes that maximally $\psi$-epistemic model cannot account for the Bell nonlocal correlations. In the `orthodox' interpretation of quantum theory, the wavefunction $\psi$ alone provides the complete description of reality which itself can be considered as $\psi$-complete ontological model. Researchers have already acknowledged that Einstein had shown incompatibility of $\psi$-complete model with locality through a simple argument at the Solvay conference \cite{Bacciagaluppi09}. In Einstein's own words \cite{Schilpp70}:

{\it ... One arrives at very implausible theoretical conceptions, if one attempts to maintain the thesis that the statistical quantum theory is in principle capable of producing a complete description of an individual physical system ...}. 

Extending this result, the authors in \cite{Harrigan09} have shown that 

{\it Any $\psi$-ontic ontological model that reproduces the quantum statistics (QSTAT) violates locality}. 

In brief, these two results together assert "$\psi$-complete and/or $\psi$-ontic $\Rightarrow$ $\neg$ locality". Our theorem can be viewed as a converse of this claim as it shows that "Nonlocality (in the sense of Bell)  $\Rightarrow$ $\neg$ maximal $\psi$-epistemicity". Here we recall the result of Ref. \cite{Leifer13}, where it has been shown that ``maximally $\psi$-epistemic $\Rightarrow$ Kochen-Specker noncontextual". Therefore Kochen-Specker contextuality \cite{Kochen67} excludes maximally $\psi$-epistemic model for quantum systems with Hilbert space dimension strictly greater that two. Similarly, our theorem shows that the Bell nonlocality rules out maximally $\psi$-epistemic ontological model. Manifestly, the scope of our theorem is broader than that in \cite{Leifer13} as it excludes maximally $\psi$-epistemic model even for a qubit system.   

So far we have shown that Bell nonlocality prohibits ontological model that is maximally $\psi$-epistemic. Naturally the question arises what about the non-maximally $\psi$-epistemic models [see Eq. (\ref{deg})]? Interestingly, we will now show that a broader class of ontological models having certain degree of epistemicity can also be excluded from a similar reasoning.

Whenever the state $\ket{+}$ is fed into the machine copying the state $\ket{0}$ and $\ket{1}$ perfectly, the resulting state being maximally entangled exhibits Bell nonlocality. Thus part of the ontic support $\Lambda_+$ must lie outside $\Lambda_{0} \cup \Lambda_{1}$, which accounts for the observed CHSH violation and consequently this will imposes bound on degree of epistemicity. For suitable choices of measurements maximally entangled state can exhibits CHSH violation up-to $2\sqrt{2}$, {\it i.e.} $\bra{\phi^+}\mathbb{CHSH}\ket{\phi^+}=2\sqrt{2}$; $\mathbb{CHSH}$ denotes the CHSH expression/operator. 
Quantum reproducibility condition, therefore, demands
\begin{align*}
&\int_{\Lambda_+} d\lambda^S \int_{\Lambda^R}\int_{\Lambda^G}d\lambda^R d\lambda^G \mu(\lambda_{joint}|\ket{+r})\mathbb{CHSH}_{\mathbb{M}[\lambda_{joint}]}\\
&~~~~~~~~~~~~~~~~~~~~=\bra{\phi^+}\mathbb{CHSH}\ket{\phi^+}=2\sqrt{2}.  
\end{align*}
Here $\lambda_{joint}\equiv(\lambda^S,\lambda^R,\lambda^G)$ is the input ontic state and $\mathbb{M}[\lambda_{joint}]$ is the output ontic state after the machine's action. The domain of integration for the variable $\lambda^S$ can be divided into three disjoint parts, {\it i.e.} $\int_{\Lambda_+}d\lambda^S=\int_{\Lambda_0 \cap \Lambda_+}d\lambda^S+\int_{\Lambda_1 \cap \Lambda_+}d\lambda^S+\int_{\Lambda_+\setminus(\Lambda_0 \cup \Lambda_1)}d\lambda^S$.

Note that whenever $\lambda^S\in\Lambda_i\cap\Lambda_+$, the joint ontic state $\lambda_{joint}=(\lambda^S,\lambda^R,\lambda^G)$ belonging in $\Lambda_{+r}$ as well as in $\Lambda_{ir}$ can not lead to the observed nonlocality since $\mathbb{M}[\lambda_{joint}]\in\Lambda_{ii}$. However, there may exist $(\lambda^S,\lambda'^R,\lambda'^G) \in \Lambda_{+r}$ s.t. $(\lambda^S,\lambda'^R,\lambda'^G) \notin \Lambda_{ir}$, where $\lambda^S \in \Lambda_i \cap \Lambda_+ $ for some $i\in\{0,1\}$. In such a case the machine can distinct the input preparation $\ket{+}$ or $\ket{i}$ by accessing $\lambda'^R$ and/or $\lambda'^G$ and the observed nonlocal behaviour can be well explained. At this point we assume that
\begin{align}\label{eq3}
\forall~\lambda^S &\in \Lambda_i \cap \Lambda_+,\nonumber\\
~~(\lambda^S,\lambda'^R,\lambda'^G) \in \Lambda_{+r} &\implies (\lambda^S,\lambda'^R,\lambda'^G) \in \Lambda_{ir}.
\end{align}
Importantly, this is a strictly weaker assumption than the ontic composition principle of PI used by PBR. To argue this first note that, for $\lambda^S \in \Lambda_+ \cap \Lambda_i$ the  Bayes' rule allows us to write $\mu(\lambda^S,\lambda^R,\lambda^G|\ket{+r}) = \mu(\lambda^S|\ket{+r})\mu(\lambda^R|\lambda^S,\ket{+r})\mu(\lambda^G|\lambda^S,\lambda^R,\ket{+r})$.~We~can also write $\mu(\lambda^S,\lambda^R,\lambda^G|\ket{ir}) = \mu(\lambda^S|\ket{ir})\mu(\lambda^R|\lambda^S,\ket{ir})$ $\mu(\lambda^G|\lambda^S,\lambda^R,\ket{ir})$. Clearly, the assumptions  
\begin{subequations}\label{eq4}
\begin{align}\label{eq4a}
\mu(\lambda^R|\lambda^S,\ket{+r}) &= \mu(\lambda^R|\lambda^S,\ket{ir}),\\
\mu(\lambda^G|\lambda^S,\lambda^R,\ket{+r} &= \mu(\lambda^G|\lambda^S,\lambda^R,\ket{ir})) 
\end{align}
\end{subequations}
will imply Eq.(\ref{eq3}). Unlike the PI assumption, Eq.(\ref{eq4}) do not prohibit correlation at the ontic level for two operationally independent preparations and hence they are weaker assumptions \cite{Self}. Furthermore, conditions in Eq.(\ref{eq4}) suffice our purpose, but they are not at all the necessary requirements. The following weaker conditions \begin{subequations}\label{eq5}
\begin{align}
\mu(\lambda^R|\lambda^S,\ket{ir}&= 0 \implies \mu(\lambda^R|\lambda^S,\ket{+r})=0,~\&\\
\mu(\lambda^G|\lambda^S,\lambda^R \ket{ir}&= 0 \implies \mu(\lambda^G|\lambda^S,\lambda^R \ket{+r})=0, 
\end{align} 
\end{subequations}
serve good for our purpose. Therefore   $\Lambda_+\setminus(\Lambda_0 \cup \Lambda_1)$ should have non-zero measure in order to explain the observed nonlocality. 

The area of the domain $\Lambda_+\setminus(\Lambda_0 \cup \Lambda_1)$ required to explain the observed quantum nonlocal imposes bound on the degree of epistemicity of the underlying ontological model. Assuming $\Omega(+,0)=\Omega(+,1):=\Omega$ it turns out that $\Omega\le2-\sqrt{2}$ \cite{Supple}. Therefore a threshold amount of onticity is required in the ontological model to incorporate the observed CHSH violation in quantum theory. Taking more general consideration of a cloning machine $\mathbb{M}_\phi$ that perfectly copies the state $\ket{\phi}$ and $\ket{\phi^\perp}$, one can obtain the following general bound $|\alpha|^2\Omega(\phi,\psi)+ |\beta|^2\Omega(\phi^{\perp},\psi)\le2-\sqrt{1+4|\alpha|^2|\beta|^2}$, where $\ket{\psi}=\alpha\ket{\phi}+\beta\ket{\phi^\perp}$ and $\alpha\neq0,1$ \cite{Supple}.

\section{Discussions} 
The PBR theorem initiated a surge in research interest  regarding the reality of quantum wavefunction \cite{Hardy13,Colbeck12,Colbeck17,Patra13,Leifer14,Lewis12,Montina12,Bub12,Banik14,Jennings15,Bandyopadhyay17}. Between the two competitive views -- $\psi$-ontic vs. $\psi$-epistemic -- it favours the former. To this claim it uses an assumption, called PI, regarding the ontology of a composite system. Establishing such a powerful doctrine about the reality of quantum state, the theorem goes through detailed scrutiny and the PI assumption gained several criticism \cite{Schlosshauer12,Emerson13,Ballentine14,Schlosshauer14}. In particular the criticism in Ref. \cite{Schlosshauer14} is quite severe. The authors there consider ontic composition assumption that is weaker than the assumption of `preparation independence' and reject vast class of deterministic hidden-variables theories, including those consistent on their targeted domain. This result puts challenge on the compositional aspect of the real/ontic states one might wish to assume through preparation independence while modeling a tensor-product quantum state. It therefore motivates renewed aspiration to establish $\psi$-ontic nature of quantum wavefunction from more rational assumption or using no such assumption at all. At this point our theorem starts contributing. Our result established that if we take maximally $\psi$-epistemic doctrine then quantum nonlocality cannot be explained in such a model. Importantly, unlike PBR, we do not consider any compositional assumption. In fact, we consider that ontic state space for product preparation can be more general than Cartesian product of their individual ontic states as they can possess composite ontic properties. In this regard, the toy model of Spekkens \cite{Spekkens07} is worth mentioning. The model is maximally $\psi$-epistemic by construction. Though it reproduces number of phenomena as observed in quantum theory, it is a Bell local model. Our theorem establishes a general result in this direction as it shows that no maximally $\psi$-epistemic model can incorporate the nonlocal behavior of quantum theory. Furthermore, we show that the observed phenomenon of quantum nonlocality excludes not only the maximally $\psi$-epistemic models, but it also imposes bound on the degree of epistemicity of the underlying models. Extent of our theorem is also broader than the $\psi$-ontology theorems in  \cite{Maroney12,Maroney12(1),Leifer13,Barrett14,Leifer14(1),Branciard14,Ringbauer15} as these results apply to quantum systems with dimensions greater than two but remain silent for qubit system. Here it is worth mentioning the theorem proved by Aaronson {\it et. al.} \cite{Aaronson13}. There the authors have considered only those psi-epistemic models in which every pair of non-orthogonal states have ontic overlap of nonzero measure. However, our theorem does not limit itself to such pairwise $\psi$-epistemic models i.e. given any two non-orthogonal states, the theorem dictates an upper bound on their ontic overlap without assuming anything about the degree of ontic overlap between all other pairs of non-orthogonal states. Our result opens up new research possibilities. In our work we have considered the CHSH inequality. It would be interesting to study whether more stringent restriction(s) on the underlying ontological models can be obtained from other classes of local realistic inequalities. It may also be interesting to study what new kind of restriction genuine quantum nonlocality would impose on the nature of quantum wavefunction.    

\begin{acknowledgments}
ADB and MB acknowledge fruitful discussions with Guruprasad Kar, Some Sankar Bhattacharya, and Tamal Guha. MB would like to thank Michael J. W. Hall for useful suggestions (through private communication). MB acknowledges research grant through INSPIRE-faculty fellowship from the Department of Science and Technology, Government of India.
\end{acknowledgments} 

\appendix
\section{Proof of Proposition 1}
Before going to the main part of the proof, let us discuss first what conclusion one can make regarding the evolution of the underlying ontic states when some measurement is performed on some operational quantum preparation. 

{\bf Single system:} Consider that a qubit is prepared in the state $\ket{0}$, the $+1$ eigenstate of $\sigma_z$. At the ontological level the system is prepared in some ontic state $\lambda\in \Lambda_0:=\{\lambda~|~\mu(\lambda|
\ket{0}>0\}\subset \Lambda$. Suppose $\sigma_x$ measurement is performed on system.  

In such a scenario, what can we say about the post-measurement ontic state (say) $\lambda^\prime$? Can we give a deterministic evolution from the pre-measurement ontic state $\lambda$ to the post-measurement ontic state $\lambda^\prime$? Answer of this question is NO.  However, we can make some conclusion regarding $\lambda^\prime$. Given that on the pre-measurement ontic state $\lambda$ the measurement $\sigma_x$ is performed, we can say that post-measurement ontic state $\lambda^\prime$ must belong to  $
(\Lambda_+ \cup \Lambda_- )$ {\it i.e.} $\Lambda_0\ni\lambda\xrightarrow{\sigma_x} \lambda' \in (\Lambda_+ \cup \Lambda_-)$. 

Suppose that on the quantum preparation $\ket{0}$ the $\sigma_z$ measurement is performed instead of the $\sigma_x$ measurement. Clearly, the operational state remains identical as a measurement does not disturb its eigenkets. In this case, can we make the conclusion that the post-measurement ontic state $\lambda^\prime$ is same as the pre-measurement ontic state $\lambda$? Again the answer is NO. However, we can make the conclusion that $\lambda^\prime$ must belong to the support $\Lambda_0$. In other words, a measurement that does not disturb the operational preparation can disturb the ontic preparation. Such a disturbance is indeterministic in general but the final ontic state remains within the support on the operational preparation, {\it i.e.} $\Lambda_0\ni\lambda\xrightarrow{\sigma_z} \lambda' \in \Lambda_0$.

Therefore, in general we can say a measurement, disturbing or non-disturbing on an operational preparation, should not assume any deterministic state-update rule for the underlying ontic states. The only thing that can be specified is the set of possible post-measurement ontic states for a given pre-measurement ontic state.

{\bf Bipartite system:} Consider a two-qubit bipartite system $\mathcal{S}_{AB}$. As discussed in the manuscript, the ontic state space of such a system has three parts: $\Lambda_A\times\Lambda_B$ corresponding local reality, $\Lambda^G$ carrying strictly relational information about the two systems and remains hidden under local measurements, and $\Lambda^{NL}$ reality resulting to the Bell nonlocal feature. Since in the following we will consider local measurements on the bipartite system, without loss of any generality we only consider the joint ontic states $\lambda_{joint}\in(\Lambda_A\times\Lambda_B)\cup\Lambda^{NL}$. Any $\lambda_{joint}\in(\Lambda_A\times\Lambda_B)$ can be thought as $\lambda_{joint}\equiv(\lambda_A,\lambda_B)$, where $\lambda_A \in \Lambda^A$ and $\lambda_B \in \Lambda^B$. On the other hand, a $\lambda_{joint}\in\Lambda^{NL}$ must violates at least one of the conditions of outcome independence (OI) and parameter independence.

{\it Outcome independence:}
\begin{subequations}
	\begin{align}
	p(a|b,x,y,\lambda_{joint})=p(a|x,y,\lambda_{joint}),~\forall~ a,b,x,y;\\
	p(b|a,x,y,\lambda_{joint})=p(b|x,y,\lambda_{joint}),~\forall~ a,b,x,y.
	\end{align}
\end{subequations}
Here, $a,b$ respectively denote Alice's and Bob's outcome for their respectively local measurements $x$ and $y$.

{\it Parameter independence:}
\begin{subequations}
	\begin{align}
	p(a|x,y,\lambda_{joint})=p(a|x,\lambda_{joint}),~\forall~ a,x,y;\label{eq2a}\\
	p(b|x,y,\lambda_{joint})=p(b|y,\lambda_{joint}),~\forall~ b,x,y.\label{eq2b}
	\end{align}
\end{subequations}
In a OI model, a nonlocal ontic state $\lambda_{joint}\in\Lambda^{NL}$ must violates the condition Parameter independence which can happen in three ways:\\
(i) Violation of Eq.(\ref{eq2a}) for some choice of $a,x,y$. Such a $\lambda_{joint}$ we will represent as $\lambda_{joint}\equiv(\lambda_{A \leftarrow B},\lambda_B)$, as Bob's measurement choice effects Alice's marginal outcome probabilities. \\  
(ii) Violation of Eq.(\ref{eq2b}); $\lambda_{joint}\equiv(\lambda_A,\lambda_{{A\rightarrow B}})$.\\
(iii) Violation of Eq.(\ref{eq2a}) and Eq.(\ref{eq2b}); $\lambda_{joint}\equiv\lambda_{{A\leftrightarrow B}}$.

For a better clarification, specific examples of these four types of $\lambda_{joint}$'s are discussed below.\\
{\bf Case (1):} $\lambda_{joint} \equiv (\lambda_A,\lambda_B)\in\Lambda_A\times\lambda_B$.
\begin{table}[h!]
\centering
\begin{tabular}{|c|c|c|c|c|c|}
			\hline
			$~~\#~~$ & $xy(\downarrow)/ab (\rightarrow)$ & $+1,+1$ & $+1,-1$ &$-1,+1$ &$-1,-1$ \\
			\hline\hline
			$1$ & $\sigma_x\otimes\sigma_x$  & $1$ & $0$ & $0$ & $0$\\
			\hline
			$2$ & $\sigma_x\otimes\sigma_y$  & $0$ & $1$ & $0$ & $0$\\
			\hline
			$3$ & $\sigma_y\otimes\sigma_x$  & $0$ & $0$ & $1$ & $0$\\
			\hline
			$4$ & $\sigma_y\otimes\sigma_y$  & $0$ & $0$ & $0$ & $1$\\
			\hline
\end{tabular}
\caption{Note that $\lambda_{joint}$ satisfies the condition of Parameter Independence. None of Alice's outcome depends on the measurement context(s) chosen by Bob, and vice versa. Therefore, observables of both A and B possess local reality (independent of any remote context chosen by other observer).}
\end{table}
Such a $\lambda_{joint}$ can be described by Cartesian product of ontic states $\lambda_A\in\Lambda_A$ and $\lambda_B\in\Lambda_B$ corresponding the following sub-tables. 
\begin{table}[h!]
\begin{tabular}{ll}
\begin{tabular}{|c|c|c|c|c|c|}
				\hline
				$~~\#~~$ & $x(\downarrow)/a (\rightarrow)$ & $+1$ & $-1$ \\
				\hline\hline
				$1$ & $\sigma_x$  & $1$ & $0$ \\
				\hline
				$2$ & $\sigma_y$  & $0$ & $1$ \\
				\hline
\end{tabular}
&~~~~~~
\begin{tabular}{|c|c|c|c|c|c|}
				\hline
				$~~\#~~$ & $y(\downarrow)/b (\rightarrow)$ & $+1$ & $-1$ \\
				\hline\hline
				$1$ & $\sigma_x$  & $1$ & $0$ \\
				\hline
				$2$ & $\sigma_y$  & $0$ & $1$ \\
				\hline
\end{tabular}
\end{tabular}
\end{table}
{\bf Case (2-i):} $\lambda_{joint} \equiv (\lambda_{A\leftarrow B},\lambda_B)\in\Lambda^{NL}$.
\begin{table}[h!]
\centering
\begin{tabular}{|c|c|c|c|c|c|}
		\hline
			$~~\#~~$ & $xy(\downarrow)/ab (\rightarrow)$ & $+1,+1$ & $+1,-1$ &$-1,+1$ &$-1,-1$ \\
			\hline\hline
			$1$ & $\sigma_x\otimes\sigma_x$  & $1$ & $0$ & $0$ & $0$\\
			\hline
			$2$ & $\sigma_x\otimes\sigma_y$  & $1$ & $0$ & $0$ & $0$\\
			\hline
			$3$ & $\sigma_y\otimes\sigma_x$  & $1$ & $0$ & $0$ & $0$\\
			\hline
			$4$ & $\sigma_y\otimes\sigma_y$  & $0$ & $0$ & $1$ & $0$\\
			\hline
\end{tabular}
\caption{Note that Alice's outcome for $\sigma_y$ measurement depends on the Bob's measurement choice(s) (compare $3^{rd}$ and $4^{th}$ rows). However, Bob's outcomes are independent of measurement contexts chosen by Alice.}
\end{table}\\
In this case, observables of A system do not possess any local reality, {\it i.e.} it cannot be described by an ontic state belonging to its local ontic space $\Lambda^A$. However, observables of B system take values independent of measurement contexts chosen by Alice, hence they possess local ontic reality, {\it i.e.} B has its local ontic state $\lambda_B \in \Lambda^B$.
\begin{table}[h!]
\begin{tabular}{ll}
\begin{tabular}{|c|c|c|c|c|c|}
				\hline
				$~~\#~~$ & $y(\downarrow)/b (\rightarrow)$ & $+1$ & $-1$ \\
				\hline\hline
				$1$ & $\sigma_x$  & $1$ & $0$ \\
				\hline
				$2$ & $\sigma_y$  & $1$ & $0$ \\
				\hline
\end{tabular}
\end{tabular}
\end{table}\\
{\bf Case(2-ii):} $\lambda_{joint} \equiv (\lambda_A,\lambda_{A\rightarrow B})\in\Lambda^{NL}$.
\begin{table}[h!]
\centering
\begin{tabular}{|c|c|c|c|c|c|}
			\hline
			$~~\#~~$ & $xy(\downarrow)/ab (\rightarrow)$ & $+1,+1$ & $+1,-1$ &$-1,+1$ &$-1,-1$ \\
			\hline\hline
			$1$ & $\sigma_x\otimes\sigma_x$  & $1$ & $0$ & $0$ & $0$\\
			\hline
			$2$ & $\sigma_x\otimes\sigma_y$  & $1$ & $0$ & $0$ & $0$\\
			\hline
			$3$ & $\sigma_y\otimes\sigma_x$  & $1$ & $0$ & $0$ & $0$\\
			\hline
			$4$ & $\sigma_y\otimes\sigma_y$  & $0$ & $1$ & $0$ & $0$\\
			\hline
\end{tabular}
\caption{Note that $\sigma_y$'s value for Bob depends on the measurement context chosen by Alice (compare $2^{nd}$ and $4^{th}$ rows). Importantly, Alice's outcomes do not depend on Bob's measurement choice(s).}
\end{table}\\
In this case outcome of the observables on A system do not depend on Bob's measurement choice. Therefore, though B does not possess local reality ( {\it i.e.} context independent reality) the system A is in a local ontic state $\lambda^A \in \Lambda^A $.
\begin{table}[h!]
\begin{tabular}{ll}
\begin{tabular}{|c|c|c|c|c|c|}
				\hline
				$~~\#~~$ & $x(\downarrow)/a (\rightarrow)$ & $+1$ & $-1$ \\
				\hline\hline
				$1$ & $\sigma_x$  & $1$ & $0$ \\
				\hline
				$2$ & $\sigma_y$  & $1$ & $0$ \\
				\hline
\end{tabular}
\end{tabular}
\end{table}\\
{\bf Case(2-iii):} $\lambda_{joint} \equiv \lambda_{A \leftrightarrow B}$.
\begin{table}[h!]
\centering
\begin{tabular}{|c|c|c|c|c|c|}
			\hline
			$~~\#~~$ & $xy(\downarrow)/ab (\rightarrow)$ & $+1,+1$ & $+1,-1$ &$-1,+1$ &$-1,-1$ \\
			\hline\hline
			$1$ & $\sigma_x\otimes\sigma_x$  & $1$ & $0$ & $0$ & $0$\\
			\hline
			$2$ & $\sigma_x\otimes\sigma_y$  & $1$ & $0$ & $0$ & $0$\\
			\hline
			$3$ & $\sigma_y\otimes\sigma_x$  & $0$ & $1$ & $0$ & $0$\\
			\hline
			$4$ & $\sigma_y\otimes\sigma_y$  & $0$ & $0$ & $0$ & $1$\\
			\hline
\end{tabular}
\caption{Parameter independence is violated both from A to B and B to A. Alice's $\sigma_y$ measurement takes different outcome depending on Bob's measurement context. Similarly, Bob's $\sigma_x$ outcome depends Alice's measurement context. Therefore, none of Alice's and Bob's observables possess local reality.}
\end{table}

We are now in a position to prove the Proposition stated in the main text. 
\begin{proof}
Consider the product quantum preparation $\ket{0}_A\ket{0}_B$. Contrary to Proposition 1, assume that there exists some nonlocal joint ontic state $\lambda_{joint}\in\Lambda^{NL}$ lying in the support $\Lambda_{00}$. Since, we are considering the ontological model to be maximally $\psi$-epistemic, therefore the model must be outcome deterministic and reciprocal \cite{Ballentine14}. Thus the nonlocal $\lambda_{AB}$ must violate at least one of the assumptions of parameter independence and outcome independence (OI) \cite{Jarrett84,Shimony84}. Since outcome deterministic models satisfy OI \cite{Hall11}, therefore $\lambda_{AB}$ must violate parameter independence. Note that, nonlocal correlations have a classical like explanation if we assume that the agent sacrifice their {\it free choice}/ measurement independence \cite{Hall10,Barrett11,Banik12,Bub12,Banik13}. However, here we are considering that the agents enjoy their free choice.
	
Consider a $\lambda_{joint}\equiv \lambda_{A \leftrightarrow B}$ [{\bf Type (2-iii)} nonlocal]. Such a $\lambda_{joint}$ requires two different tables of assignments: one for when Alice measures first (which would be [$\bf Type (2-ii)$ non-local]), and a different one when Bob measures first (which would be [$\bf Type (2-i) $] local). But for simplicity, we consider a fixed order of their measurements : Alice measures her part of system before Bob performs any measurement on his part , so that a single table of assignments is sufficient. A typical example of such $\lambda_{joint}  \equiv \lambda_{A \leftrightarrow B} \in \Lambda_{00}$ is  the following : 
\begin{table}[h!]
\centering
\begin{tabular}{|c|c|c|c|c|c|}
				\hline
				$~~\#~~$ & $xy(\downarrow)/ab (\rightarrow)$ & $+1,+1$ & $+1,-1$ &$-1,+1$ &$-1,-1$ \\
				\hline\hline
				$1$ & $\sigma_z\otimes\sigma_z$  & $1$ & $0$ & $0$ & $0$\\
				\hline
				$2$ & $\sigma_z\otimes\sigma_x$  & $1$ & $0$ & $0$ & $0$\\
				\hline
				$3$ & $\sigma_x\otimes\sigma_z$  & $1$ & $0$ & $0$ & $0$\\
				\hline
				$4$ & $\sigma_x\otimes\sigma_x$  & $1$ & $0$ & $0$ & $0$\\
				\hline
				$5$ & $\sigma_x\otimes\sigma_y$  & $1$ & $0$ & $0$ & $0$\\
				\hline
				$6$ & $\sigma_y\otimes\sigma_x$  & $0$ & $1$ & $0$ & $0$\\
				\hline
				$7$ & $\sigma_y\otimes\sigma_y$  & $1$ & $0$ & $0$ & $0$\\
				\hline
				$8$ & $\sigma_z\otimes\sigma_y$  & $1$ & $0$ & $0$ & $0$\\
				\hline
				$9$ & $\sigma_y\otimes\sigma_z$  & $0$ & $1$ & $0$ & $0$\\
				\hline
				$\vdots$ & $\vdots$  & $\vdots$ & $\vdots$ & $\vdots$ & $\vdots$\\
				\hline
\end{tabular}
\caption{$\lambda_{joint}  \equiv \lambda_{A \leftrightarrow B}$.}\label{tab5}
\end{table}

Note that in accordance with our chosen order of measurements, the table of assignment for $\lambda_{joint} \equiv \lambda_{A\leftrightarrow B}$ is of [$\bf{Type (2-ii)}$ nonlocal]. Here, we do not provide any table of [$\bf{Type (2 - i)}$ nonlocal] associated with $\lambda_{joint} \equiv \lambda_{A\leftrightarrow B}$ that considers the case when Alice and Bob measures in the reverse temporal order (i.e. Bob first, Alice second) since it is straightforward to construct such a table.
	
On this pre-measurement joint ontic state let Alice first measures $\sigma_z$ on her system. This will cause all observable at Bob's end to take values in accordance with the chosen context of $\sigma_z$ at Alice's side. Therefore, post-measurement $\lambda'_{joint}$ will be of {\bf Type (2-ii)} nonlocal, {\it i.e.} $(\lambda^\prime_{A\leftarrow B},\lambda^\prime_B)$; where $\lambda^\prime_B$ is described as the the following table: 
\begin{table}[h!]
\centering
\begin{tabular}{|c|c|c|c|c|c|}
			\hline
			$~~\#~~$ & $y(\downarrow)/b (\rightarrow)$ & $+1$ & $-1$  \\
			\hline\hline
			$1$ & $\sigma_z$  & $1$ & $0$ \\
			\hline
			$2$ & $\sigma_x$  & $1$ & $0$ \\
			\hline
			$3$ & $\sigma_y$  & $1$ & $0$ \\
			\hline
			$\vdots$ & $\vdots$  & $\vdots$ & $\vdots$ \\
			\hline
\end{tabular}
\caption{}\label{tab6}
\end{table}

The measurement $\sigma_z$ on A system thus leads to the following ontic transformation:
\begin{align}
\lambda^{joint}&\equiv\lambda_{A \leftrightarrow B} \xrightarrow[]{\sigma_z \otimes \mathbb{I}} \lambda^\prime_{joint}\nonumber\\
&\equiv(\lambda^\prime_{A\leftarrow B},\lambda^\prime_B) \in\mathbf{Type (2-ii)}.  
\end{align}
Following reasoning clarifies such a post-measurement evolution of the ontic state.  Before Alice's measurement, observables of system B possess no local reality, which is reflected in the pre-measurement ontic state $\lambda_{A \leftrightarrow B}$. For instance, Bob's $\sigma_x$ observable does not have context independent reality in the state $\lambda_{A \leftrightarrow B}$ (see Table \ref{tab5}). In other words, one cannot assign value to $\sigma_x$ of B independent of Alice's choice of measurement (From the $2^{nd}$ and $6^{th}$ row of Table \ref{tab5} it is evident.) But, as soon as Alice chooses one particular measurement context Bob's observables take the corresponding values in accordance with the Alice's chosen measurement context. To see that consider the $2^{nd}$ row of pre-measurement state (Table \ref{tab5}) which asserts: ``\textit{If Alice measures $\sigma_z$ on A, Bob will obtain +1 if he measures $\sigma_x$ on B}". When Alice already performed the measurement $\sigma_z$, Bob's $\sigma_x$ observable must take the value $+1$ in the post-measurement state $\lambda'_{joint}$, otherwise proposition of the $2^{nd}$ row of pre-measurement state gets violated. If Alice had measured $\sigma_y$ instead of $\sigma_z$, Bob's $\sigma_x$ observable would have taken the value $-1$, otherwise proposition of the $6^{th}$ row of pre-measurement state would have been violated. Since Alice measures $\sigma_z$, not $\sigma_y$, the possibility of Bob's $\sigma_x$ value to be $-1$ is no more there in the post measurement state $\lambda'_{joint}$. Therefore Alice's choice of a particular context assigns Bob's observable context independent local reality. Similarly, after Alice's $\sigma_z$ measurement, $8^{th}$ row demands B should have $\sigma_y = 1$ in the post-measurement joint ontic state $\lambda'_{joint}$.
	
In general, in the post-measurement joint ontic state $\lambda'_{joint}$  ,  $\sigma_n$ of B will take the value obeying the condition imposed by the row corresponding to $\sigma_z \otimes \sigma_n$ of pre-measurement joint ontic state $\lambda_{A\leftrightarrow B}$. Thus all the observables of B take some fixed particular value, therefore attain local reality, as soon as Alice chooses her measurement context. Therefore, in the post-measurement state, though Parameter Independence can be violated from B to A there is no such violation from A to B and hence in $\lambda'_{joint}$, B is in the ontic state $\lambda^\prime_B\in\Lambda_B$ (given in Table \ref{tab6}).
	
Please note that we can only claim that the post measurement ontic state $\lambda_{joint}\equiv(\lambda^\prime_{A\leftarrow B},\lambda^\prime_B) \in\mathbf{Type (2-ii)}$, but which particular $\lambda'_{joint}$ it is that is not specified and for our argument it is not required. 	Suppose Bob now performs $\sigma_z$ measurement on the state $\lambda^\prime_{joint}$. A similar reasoning as above will imply the final ontic state $\lambda^{\prime\prime}_{joint}\equiv(\lambda^{\prime\prime}_{A},\lambda^{\prime\prime}_{B})\in\mathbf{Type (1)}$. Thus we have,
\begin{align*}
\mathbf{Type (2-iii)}\ni\lambda_{A \leftrightarrow B} \xrightarrow[]{\sigma_z \otimes \mathbb{I}} (\lambda^\prime_{A\leftarrow B},\lambda^\prime_B)\in\mathbf{Type (2-ii)}\nonumber\\
\xrightarrow[]{\mathbb{I} \otimes\sigma_z}(\lambda^{\prime\prime}_{A},\lambda^{\prime\prime}_{B})\in\mathbf{Type (1)}.  
\end{align*}
Here, in the second step of ontic evolution in equation $(4)$, we have made the following {\it assumption regarding ontic evolution} of a certain class of joint ontic states.
	
In any joint ontic state $\lambda_{joint}^{AB}$, if there exists no parameter dependence from $A \rightarrow B$ (or from $B \rightarrow A$), local measurement on B ( or A) cannot generate parameter dependence from $A \rightarrow B$ (or $B \rightarrow A$ ) in the post-measurement state $\lambda^{\prime AB}_{joint}$. Due to this, in the second step of ontic evolution in $(4)$, the following has not been the case : $(\lambda^\prime_{A\leftarrow B},\lambda^\prime_B)\in\mathbf{Type (2-ii)} \xrightarrow[]{\mathbb{I} \otimes\sigma_z}(\lambda^{\prime\prime}_{A},\lambda^{\prime\prime}_{A\rightarrow B})\in \mathbf{Type (2-i)}.  $ More precisely, as there was no parameter dependence from $A\rightarrow B$  in $(\lambda^{\prime}_{A\leftarrow B},\lambda^{\prime}_{B})$, measuring $\sigma_z$ on B cannot generate parameter dependence from $A\rightarrow B$ in the post-measurement state.
	
As the local joint ontic states [ i.e. $\lambda_{joint} \equiv (\lambda_A, \lambda_B) \in \bf{Type(1)}$ ] do not possess parameter dependence in either direction ( i.e. neither $A\rightarrow B$ nor $A\leftarrow B$), the assumption immediately leads to the following:
\begin{align*}
(\lambda_A,\lambda_B) \in  \mathbf{Type (1)} \xrightarrow[]{\sigma_z \otimes \mathbb{I}} (\lambda'_A,\lambda'_B) \in  \mathbf{Type (1)} \\ \xrightarrow[]{\mathbb{I} \otimes \sigma_z} (\lambda''_A,\lambda''_B) \in  \mathbf{Type (1)}.
\end{align*}
Joint ontic states of system AB which are local remain local under local measurements done on subsystems A and/or B. [Note : as we have fixed a specific temporal order in which Alice measures first and Bob measures second, $\lambda_{A\leftrightarrow B}$ $\in \mathbf{Type (2-iii)}$ is essentially of $\mathbf{Type (2-ii)}$ in which no parameter dependence exists from $B\rightarrow A$. Therefore, we could have as well used the assumption in the very first step of ontic evolution in $(4)$ to obtain $\mathbf{Type (2-iii)}\ni\lambda_{A \leftrightarrow B} \xrightarrow[]{\sigma_z \otimes \mathbb{I}} (\lambda^\prime_{A},\lambda^\prime_B)\in\mathbf{Type (1)}$]. 

Therefore from eq $(4)$, for any pre-measurement $\lambda_{joint}\in\Lambda^{NL}$ that were assumed to lie within the ontic support of $\ket{00}$, the post-measurement joint ontic state after Alice and Bob performing their local measurements will be $\lambda_{joint}\equiv (\lambda^{\prime\prime}_{A},\lambda^{\prime\prime}_{B})\in\Lambda_A\times\Lambda_B$. Therefore the post-measurement quantum state cannot contain any $\lambda_{joint} \in \Lambda^{NL}$ in its ontic support. But since the post measurement quantum state remains unchanged, {\it i.e.} it is again $\ket{0}_A\ket{0}_B$; this implies ontic support of $\ket{0}_A\ket{0}_B$ cannot contain any $\lambda_{joint}\in\Lambda^{NL}$, which is in contradiction with our initial assumption. 

It is not hard to see that one arrives at similar contradiction for every product quantum preparation. Furthermore, the arguments also holds in any outcome deterministic ontological model instead of the maximally $\psi$-epistemic one only. This completes our proof. 
\end{proof}
\section{General bound on degree of epistemicity}
\begin{figure}[t!]
\begin{center}
	\includegraphics[height=3.5cm, width=8cm]{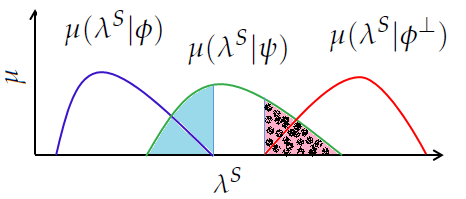}
\end{center}
\caption{(Color on-line) Ontic support $\Lambda_{\chi}\equiv\mbox{Supp}[\mu(\lambda|\chi)]$ for $\ket{\chi}\in\left\{\ket{\phi},\ket{\phi^\perp},\ket{\psi}=\alpha\ket{\phi}+\beta\ket{\phi^\perp}\right\}$. Along $x$-axis ontic states for the system $S$ are shown and along $y$-axis probability distribution on these ontic states for different quantum preparations are plotted. In case of a non-maximally $\psi$-epistemic model, $\Lambda_\psi\cap(\Lambda_{\phi}\cup\Lambda_{\phi^\perp})^{\mathrm{C}}\neq\emptyset$. Area of the sky-blue shaded region is given by $\int_{\Lambda_{\phi}}d\lambda^S\mu(\lambda^S|\psi)$ and area of the pink shaded (black dotted) region is $\int_{\Lambda_{\phi^\perp}}d\lambda^S\mu(\lambda^S|\psi)$. }\label{fig1}
\label{fig} 
\end{figure}

Consider a quantum copying machine $\mathbb{M}_\phi$ that perfectly copies the state $\ket{\phi}$ and $\ket{\phi^\perp}$. While the machine is fed with system $S$ prepared in state $\ket{\phi}$ and $\ket{\phi^\perp}$ and the reference system $R$ prepared in some fixed state $\ket{r}$, its action at the ontological level is similar as discussed in the manuscript. If we fed $\mathbb{M}_\phi$ with $\ket{\psi}_S\ket{r}_{R}$, with $\ket{\psi} = \alpha \ket{ \phi} + \beta\ket{\phi^\perp}~(\alpha\neq 0,1)$  then linearity of quantum theory results in the output state $\ket{\Theta}_{SR}= \alpha\ket{\phi}_S\ket{\phi}_R + \beta \ket{{\phi^\perp}}_S\ket{\phi^{\perp}}_R$. For suitable choices of measurements this state can exhibits CHSH inequality violation up-to $2\sqrt{1+4|\alpha|^2|\beta|^2}$ \cite{Gisin91}. Assuming maximally $\psi$-epistemicity an argument similar to the one presented in the manuscript implies $\mbox{Supp}[\mu(\lambda_{joint}|U_{\mathbb{M}_\phi}[\ket{\psi r}])]\cap\Lambda^{NL}=\emptyset$ and hence the observed nonlocality cannot be explained. To incorporate this nonlocality we need to depart from $\psi$-maximal epistemicity. Now quantum reproducibility condition for the observed nonlocality demands,
\footnotesize
\begin{align}
&\int_{\Lambda^S} d\lambda^S  \int_{\Lambda^R} \int_{\Lambda^G}d\lambda^R d\lambda^G \mu(\lambda_{joint}|\ket{\psi r})\mathbb{CHSH}_{\mathbb{M}_\phi[\lambda_{joint}]}\nonumber\\
&= \bra{\Theta_{SR}}\mathbb{CHSH}\ket{\Theta_{SR}}= 2\sqrt{1+4|\alpha|^2|\beta|^2}. \label{eq1}
\end{align}
\normalsize
where, $\lambda_{joint}=(\lambda^S, \lambda^R,\lambda^G)$ and $\mathbb{M}_\phi[\lambda_{joint}]$ is the evolved ontic state after the machine action, and $\mathbb{CHSH}$ denotes the CHSH expression. As argued in the manuscript $\mathbb{M}_\phi[\lambda_{joint}]\notin\Lambda^{NL}$ whenever $\lambda_{joint}\in(\Lambda_{\psi}\cap \Lambda_\phi)\cup(\Lambda_{\psi}\cap \Lambda_{\phi^\perp})$. Let us, therefore, consider that the ontic region $\Lambda_{\psi}\setminus(\Lambda_{\phi}\cup \Lambda_{\phi^\perp})$ has nonzero measure, which will be the case for non-maximally $\psi$-epistemic models. Thus, the domain of integration for $\lambda^S$, in the left hand side of Eq. (\ref{eq1}) can be divided into three disjoint regions $\Lambda_{\psi}\cap \Lambda_\phi$, $\Lambda_{\psi}\cap \Lambda_{\phi^\perp}$ and $\Lambda_\psi\setminus (\Lambda_{\phi}\cup \Lambda_{\phi^\perp})$ [see Fig. \ref{fig}]. Thus we have
\footnotesize
\begin{align}
&\int_{\Lambda_{\psi}\cap\Lambda_{\phi}} d\lambda^S \int d\lambda^R d\lambda^G \mu(\lambda^S,\lambda^R,\lambda^G|\ket{\psi r}) \mathbb{CHSH}_{\mathbb{M}_\phi[\lambda_{joint}]}\nonumber\\
&=\int_{\Lambda_{\psi}\cap\Lambda_{\phi}} d\lambda^S \int d\lambda^R d\lambda^G \mu(\lambda^S,\lambda^R,\lambda^G|\ket{\psi r})\times 2\nonumber\\  
&=2\int_{\Lambda_{\phi}} d\lambda^S\mu(\lambda^S|\psi)= 2 \Omega(\phi,\psi)|\langle\phi|{\psi}\rangle|^2\nonumber\\  
&=2|\alpha|^2 \Omega(\phi,\psi). \label{eq2} 
\end{align}
\normalsize
We have consider that a local ontic states yields the maximum possible value $2$ for the CHSH expression, and used the fact that $\int d\lambda^R d\lambda^G \mu(\lambda^S,\lambda^R,\lambda^G|\ket{\psi r})=\mu(\lambda^S|\psi r)=\mu(\lambda^S|\psi)$. Similar reasoning yields 
\footnotesize
\begin{align}
&\int_{\Lambda_{\psi}\cap\Lambda_{\phi^\perp}} d\lambda^S \int d\lambda^R d\lambda^G \mu(\lambda^S,\lambda^R,\lambda^G|\ket{\psi r}) \mathbb{CHSH}_{\mathbb{M}_\phi[\lambda_{joint}]}\nonumber\\
&=2|\beta|^2 \Omega(\phi^\perp,\psi). \label{eq3}    
\end{align}
\normalsize
Whenever, $\lambda_{joint}\in\Lambda_\psi\setminus (\Lambda_{\phi}\cup \Lambda_{\phi^\perp})$, the evolved ontic state $\mathbb{M}_\phi[\lambda_{joint}]$ may lie in $\Lambda^{NL}$ and consequently contributes to the observed quantum nonlocality. Assuming that all such $\lambda_{joint}$'s yield the maximum possible CHSH value ({\it i.e.} $4$) we obtain
\footnotesize
\begin{align}
&\int_{\Lambda_{\psi}\setminus(\Lambda_{\phi}\cup\Lambda_{\phi^\perp})} d\lambda^S \int d\lambda^R d\lambda^G \mu(\lambda^S,\lambda^R,\lambda^G|\ket{\psi r}) \mathbb{CHSH}_{\mathbb{M}_\phi[\lambda_{joint}]}\nonumber\\
&~=\int_{\Lambda_{\psi}\setminus(\Lambda_{\phi}\cup\Lambda_{\phi^\perp})} d\lambda^S \int d\lambda^R d\lambda^G \mu(\lambda^S,\lambda^R,\lambda^G|\ket{\psi r})\times 4\nonumber\\
&=4\left[\int_{\Lambda_{\psi}} d\lambda^S\mu(\lambda^S|\psi)-\int_{\Lambda_{\phi}} d\lambda^S\mu(\lambda^S|\psi)-\int_{\Lambda_{\phi^\perp}} d\lambda^S\mu(\lambda^S|\psi)\right]\nonumber\\
&=4\left[1-|\alpha|^2\Omega(\phi,\psi) - |\beta|^2\Omega(\phi^{\perp},\psi)\right].\label{eq4}
\end{align}
\normalsize
Eqs. (\ref{eq1}), (\ref{eq2}), (\ref{eq3}), \& (\ref{eq4}) all-together therefore imply
\footnotesize
\begin{align}
2\sqrt{1+4|\alpha|^2|\beta|^2}&\le 2~|\alpha|^2\Omega(\phi,\psi)+2~|\beta|^2\Omega(\phi^{\perp},\psi)~\nonumber\\
&~~~~~~+ 4~\left[1-|\alpha|^2\Omega(\phi,\psi) - |\beta|^2\Omega(\phi^{\perp},\psi)\right].
\end{align}
\normalsize
Here, instead of equality, we put inequality as some local $\lambda$'s can yield CHSH value less than $2$ and some nonlocal $\lambda$'s can yield CHSH value less than $4$. Simplifying the above expression we obtain  
\begin{equation}
|\alpha|^2\Omega(\phi,\psi)+ |\beta|^2\Omega(\phi^{\perp},\psi)\le2-\sqrt{1+4|\alpha|^2|\beta|^2}.   
\end{equation}


\begin{thebibliography}{99}

\bibitem{Bacciagaluppi09} G. Bacciagaluppi and A. Valentini; Quantum Theory at the Crossroads: Reconsidering the 1927 Solvay Conference; Cambridge University Press (2009)  [\href{https://arxiv.org/abs/quant-ph/0609184}{arXiv:quant-ph/0609184}].

\bibitem{Harrigan09} N. Harrigan and R. W. Spekkens; Einstein, incompleteness, and the epistemic view of quantum states; \href{https://doi.org/10.1007/s10701-009-9347-0}{Found Phys {P\bf40}, 125 (2010)}.

\bibitem{Spekkens05} R. W. Spekkens; Contextuality for preparations, transformations, and unsharp measurements; \href{https://doi.org/10.1103/PhysRevA.71.052108}{Phys. Rev. A {\bf 71}, 052108 (2005)}.

\bibitem{Harrigan07} N. Harrigan and T. Rudolph; Ontological models and the interpretation of contextuality; \href{https://arxiv.org/abs/0709.4266}{arXiv:0709.4266 [quant-ph]}.

\bibitem{Pusey12} M. F. Pusey, J. Barrett, and T. Rudolph; On the reality of the quantum state; \href{https://doi.org/10.1038/nphys2309}{Nature Phys. {\bf 8}, 475 (2012)}

\bibitem{Hardy13} L. Hardy; Are Quantum States Real? \href{https://doi.org/10.1142/S0217979213450124}{International Journal of Modern Physics B {\bf 27}, 1345012 (2013)}.

\bibitem{Colbeck12} R. Colbeck and R. Renner; Is a System’s Wave Function in One-to-One Correspondence with Its Elements of Reality? \href{https://doi.org/10.1103/PhysRevLett.108.150402}{Phys. Rev. Lett. {\bf 108}, 150402 (2012)}.

\bibitem{Colbeck17} R. Colbeck and R. Renner; A system's wave function is uniquely determined by its underlying physical state; \href{https://iopscience.iop.org/article/10.1088/1367-2630/aa515c}{New J. Phys {\bf 19}, 013016 (2017)}.

\bibitem{Patra13} M. K. Patra, S. Pironio, and S. Massar; No-Go Theorems for $\psi$-Epistemic Models Based on a Continuity Assumption; \href{https://doi.org/10.1103/PhysRevLett.111.090402}{Phys. Rev. Lett. {\bf 111}, 090402 (2013)}

\bibitem{Leifer14} M. S. Leifer; Is the quantum state real? An extended review of $\psi$-ontology theorems; \href{https://doi.org/10.12743/quanta.v3i1.22}{Quanta {\bf 3}, 67 (2014)}.

\bibitem{Schlosshauer12} M. Schlosshauer and A. Fine; Implications of the Pusey-Barrett-Rudolph Quantum No-Go Theorem; \href{https://doi.org/10.1103/PhysRevLett.108.260404}{Phys. Rev. Lett. {\bf 108}, 260404 (2012)}.

\bibitem{Emerson13} J. Emerson, D. Serbin, C. Sutherland, and V. Veitch; The whole is greater than the sum of the parts: on the possibility of purely statistical interpretations of quantum theory; \href{https://arxiv.org/abs/1312.1345}{arXiv:1312.1345 [quant-ph]}.

\bibitem{Ballentine14} L. Ballentine; Ontological Models in Quantum Mechanics: What do they tell us? \href{https://arxiv.org/abs/1402.5689}{	arXiv:1402.5689 [quant-ph]}.

\bibitem{Schlosshauer14} M. Schlosshauer and A. Fine; No-Go Theorem for the Composition of Quantum Systems; \href{https://doi.org/10.1103/PhysRevLett.112.070407}{Phys. Rev. Lett. {\bf 112}, 070407 (2014)}.

\bibitem{Maroney12} O. J. E. Maroney; How statistical are quantum states? \href{https://arxiv.org/abs/1207.6906}{arXiv:1207.6906 [quant-ph]}.

\bibitem{Maroney12(1)} O. J. E. Maroney; A brief note on epistemic interpretations and the Kochen-Specker theorem; \href{https://arxiv.org/abs/1207.7192}{arXiv:1207.7192 [quant-ph]}.

\bibitem{Leifer13} M. S. Leifer and O. J. E. Maroney; Maximally Epistemic Interpretations of the Quantum State and Contextuality; \href{https://journals.aps.org/prl/abstract/10.1103/PhysRevLett.110.120401}{Phys. Rev. Lett. {\bf 110}, 120401 (2013)}.

\bibitem{Barrett14} J. Barrett, E. G. Cavalcanti, R. Lal, and O. J. E. Maroney; No $\psi$-Epistemic Model Can Fully Explain the Indistinguishability of Quantum States; \href{https://doi.org/10.1103/PhysRevLett.112.250403}{Phys. Rev. Lett. {\bf 112}, 250403 (2014)}.

\bibitem{Leifer14(1)} M. S. Leifer; $\psi$-Epistemic Models are Exponentially Bad at Explaining the Distinguishability of Quantum States; \href{https://doi.org/10.1103/PhysRevLett.112.160404}{Phys. Rev. Lett. {\bf 112}, 160404 (2014)}. 

\bibitem{Branciard14} C. Branciard; How $\psi$-Epistemic Models Fail at Explaining the Indistinguishability of Quantum States; \href{https://doi.org/10.1103/PhysRevLett.113.020409}{Phys. Rev. Lett. {\bf 113}, 020409 (2014)}.

\bibitem{Ringbauer15} M. Ringbauer, B. Duffus, C. Branciard, E. G. Cavalcanti, A. G. White, and A. Fedrizzi; Measurements on the reality of the wavefunction; \href{https://doi.org/10.1038/nphys3233}{Nature Phys {\bf 11}, 249 (2015)}.

\bibitem{Bell64} J. S. Bell; On the Einstein Podolsky Rosen paradox; \href{https://doi.org/10.1103/PhysicsPhysiqueFizika.1.195}{Physics {\bf 1}, 195 (1964)}; Reprinted in J. S. Bell; Speakable and Unspeakable In Quantum Mechanics, (Cambridge University Press, England, Cambridge, 2004)

\bibitem{Bell66} J. S. Bell; On the Problem of Hidden Variables in Quantum Mechanics; \href{https://doi.org/10.1103/RevModPhys.38.447}{Rev. Mod. Phys. {\bf 38}, 447 (1966)}.

\bibitem{Mermin93} N. D. Mermin; Hidden variables and the two theorems of John Bell; \href{https://doi.org/10.1103/RevModPhys.65.803}{Rev. Mod. Phys. {\bf 65}, 803 (1993)}.

\bibitem{Brunner14} N. Brunner, D. Cavalcanti, S. Pironio, V. Scarani, and S. Wehner; Bell nonlocality; \href{https://doi.org/10.1103/RevModPhys.86.419}{Rev. Mod. Phys. {\bf 86}, 419 (2014)}.

\bibitem{Werner89} R. F. Werner; Quantum states with Einstein-Podolsky-Rosen correlations admitting a hidden-variable model; \href{https://doi.org/10.1103/PhysRevA.40.4277}{Phys. Rev. A {\bf 40}, 4277 (1989)}.

\bibitem{Barrett02} J. Barrett; Nonsequential positive-operator-valued measurements on entangled mixed states do not always violate a Bell inequality; \href{https://doi.org/10.1103/PhysRevA.65.042302}{Phys. Rev. A {\bf 65}, 042302 (2002)}.

\bibitem{Bowles15} J. Bowles, F. Hirsch, M. T. Quintino, and N. Brunner; Local Hidden Variable Models for Entangled Quantum States Using Finite Shared Randomness; \href{https://doi.org/10.1103/PhysRevLett.114.120401}{Phys. Rev. Lett. {\bf 114}, 120401 (2015)}.


\bibitem{Bennett99} C. H. Bennett, D. P. DiVincenzo, C. A. Fuchs, T. Mor, E. Rains, P. W. Shor, J. A. Smolin, and W. K. Wootters; Quantum nonlocality without entanglement; \href{https://doi.org/10.1103/PhysRevA.59.1070}{Phys. Rev. A {\bf 59}, 1070 (1999)}.

\bibitem{Halder19} S. Halder, M. Banik, S. Agrawal, and S. Bandyopadhyay; Strong Quantum Nonlocality without Entanglement; \href{https://doi.org/10.1103/PhysRevLett.122.040403}{Phys. Rev. Lett. {\bf 122}, 040403 (2019)}.

\bibitem{Bhattacharya20} S. S. Bhattacharya, S. Saha, T. Guha, and M. Banik; Nonlocality Without Entanglement: Quantum Theory and Beyond; \href{https://doi.org/10.1103/PhysRevResearch.2.012068}{Phys. Rev. Research {\bf 2}, 012068(R) (2020)}.

\bibitem{Supple} See the supplementary material.

\bibitem{Ionicioiu11} R. Ionicioiu and D. R. Terno; Proposal for a Quantum Delayed-Choice Experiment, \href{https://doi.org/10.1103/PhysRevLett.107.230406}{Phys. Rev. Lett. {\bf 107}, 230406 (2011)}.

\bibitem{Wootters82} W. Wootters and W. Zurek; A Single Quantum Cannot be Cloned; \href{https://doi.org/10.1038/299802a0}{Nature {\bf 299}, 802 (1982)}.

\bibitem{Clauser69} J. F. Clauser, M. A. Horne, A. Shimony, and R. A. Holt; Proposed Experiment to Test Local Hidden-Variable Theories; \href{https://doi.org/10.1103/PhysRevLett.23.880}{Phys. Rev. Lett. {\bf 23}, 880 (1969)}.

\bibitem{Gisin91} N. Gisin; Bell's inequality holds for all non-product states; \href{https://doi.org/10.1016/0375-9601(91)90805-I}{Phys. Lett. A 154, 201 (1991)}.

\bibitem{Schilpp70} P. A. Schilpp; Albert Einstein, Philosopher-Scientist: The Library of Living Philosophers (Volume VII), page 671 (1970). 

\bibitem{Kochen67} S. Kochen and E. P. Specker; The Problem of Hidden Variables in Quantum Mechanics; \href{http://dx.doi.org/10.1512/iumj.1968.17.17004}{J. Math. Mech {\bf 17}, 59 (1967)}.

\bibitem{Self}PI would demand $\mu(\lambda^R|\lambda^S,\ket{+r})= \mu(\lambda^R|\ket{r})$ and $\mu(\lambda^R|\lambda^S,\ket{ir})= \mu(\lambda^R|\ket{r})$ which implies Eq. (\ref{eq4a}), but not the converse i.e. Eq. (\ref{eq4a}) does not imply PI.

\bibitem{Lewis12} P. G. Lewis, D. Jennings, J. Barrett, and T. Rudolph; Distinct Quantum States Can Be Compatible with a Single State of Reality; \href{https://doi.org/10.1103/PhysRevLett.109.150404}{Phys. Rev. Lett. {\bf 109}, 150404 (2012)}.

\bibitem{Montina12} A. Montina; Epistemic View of Quantum States and Communication Complexity of Quantum Channels; \href{https://doi.org/10.1103/PhysRevLett.109.110501}{Phys. Rev. Lett. {\bf 109}, 110501 (2012)}.

\bibitem{Bub12} J. Bub; Bananaworld: Quantum Mechanics for Primates; \href{https://arxiv.org/abs/1211.3062}{arXiv:1211.3062 [quant-ph]}.

\bibitem{Banik14} M. Banik, S. S. Bhattacharya, S. K. Choudhary, A. Mukherjee, and A. Roy; Ontological Models, Preparation Contextuality and Nonlocality; \href{https://doi.org/10.1007/s10701-014-9839-4}{Found Phys {\bf 44}, 1230 (2014)}.

\bibitem{Jennings15} D. Jennings and M. Leifer; No return to classical reality; \href{https://doi.org/10.1080/00107514.2015.1063233}{Contemporary Physics {\bf 57}, 60 (2015)}.

\bibitem{Bandyopadhyay17} S. Bandyopadhyay, M. Banik, S. S. Bhattacharya, S. Ghosh, G. Kar, A. Mukherjee, and A. Roy; Reciprocal Ontological Models Show Indeterminism Comparable to Quantum Theory; \href{https://doi.org/10.1007/s10701-016-0058-z}{Found Phys {\bf 47}, 265 (2017)}.
 
\bibitem{Spekkens07} R. W. Spekkens; Evidence for the epistemic view of quantum states: A toy theory; \href{https://doi.org/10.1103/PhysRevA.75.032110}{Phys. Rev. A {\bf 75}, 032110 (2007)}.

\bibitem{Aaronson13} S. Aaronson, A. Bouland, L.Chua, and G. Lowther; $\psi$-epistemic theories: The role of symmetry, 
\href{https://doi.org/10.1103/PhysRevA.88.032111}{Phys. Rev. A {\bf 88}, 032111 (2013)}.

\bibitem{Jarrett84} Jon P. Jarrett; On the Physical Significance of the Locality Conditions in the Bell Arguments, \href{https://doi.org/10.2307/2214878}{No{\^{u}}s {\bf 18}, 569 (1984)}.

\bibitem{Shimony84} A. Shimony; Controllable and Uncontrollable Non-Locality, in Kamefuchi et al., eds., Foundations of Quantum Mechanics in the Light of New Technology, Tokyo, Physical Society of Japan, 225-230 (1984).

\bibitem{Hall11} Michael J. W. Hall; Relaxed Bell inequalities and Kochen-Specker theorems, \href{https://doi.org/10.1103/PhysRevA.84.022102}{Phys. Rev. A {\bf 84}, 022102 (2011)}.

\bibitem{Hall10} Michael J. W. Hall; Local Deterministic Model of Singlet State Correlations Based on Relaxing Measurement Independence, \href{https://doi.org/10.1103/PhysRevLett.105.250404}{Phys. Rev. Lett. {\bf 105}, 250404 (2010)}.

\bibitem{Barrett11} J. Barrett and N. Gisin; How Much Measurement Independence Is Needed to Demonstrate Nonlocality? \href{https://doi.org/10.1103/PhysRevLett.106.100406}{Phys. Rev. Lett. {\bf 106}, 100406 (2011)}.

\bibitem{Banik12} M. Banik, MD R. Gazi, S. Das, A. Rai, and S. Kunkri; Optimal free-will on one side in reproducing the singlet correlation, \href{https://doi.org/10.1088/1751-8113/45/20/205301}{J. Phys. A: Math. Theor. {\bf 45}, 205301 (2012)}.


\bibitem{Banik13} M. Banik; Lack of measurement independence can simulate quantum correlations even when signaling can not, \href{https://doi.org/10.1103/PhysRevA.88.032118}{Phys. Rev. A {\bf 88}, 032118 (2013)}. 

\bibitem{Gisin91} N. Gisin; Bell's inequality holds for all non-product states; \href{https://doi.org/10.1016/0375-9601(91)90805-I}{Phys. Lett. A 154, 201 (1991)}.




\end{thebibliography}
\end{document}